\documentclass[conference, a4paper]{IEEEtran}
\IEEEoverridecommandlockouts
\usepackage{cite}
\usepackage{amsmath,amssymb,amsfonts,mathtools}
\usepackage{bbm}
\usepackage{algorithmic}
\usepackage{graphicx}
\usepackage{textcomp}
\usepackage{xcolor}
\usepackage{subcaption}
\usepackage{siunitx}
\def\BibTeX{{\rm B\kern-.05em{\sc i\kern-.025em b}\kern-.08em
    T\kern-.1667em\lower.7ex\hbox{E}\kern-.125emX}}
    
\definecolor{rick}{RGB}{0,139,0}

\definecolor{Rafael}{RGB}{139,0,0}

\definecolor{Muah}{RGB}{0,0,139}

\begin{document}

\title{Learning End-to-End Channel Coding with Diffusion Models\\
\thanks{This work was supported in part by the German Federal Ministry of Education and Research (BMBF) under Grants 16KISK001K and 16KIS1004 and in part by the German Research Foundation (DFG) under the Grant SCHA 1944/7-1.}	
}

\author{
    \IEEEauthorblockN{Muah Kim, Rick Fritschek, and Rafael F. Schaefer\\}
    \IEEEauthorblockA{\centering
    \begin{tabular}{c}
    \textit{Chair of Information Theory and Machine Learning}\\
    \textit{Technical University of Dresden}\\
      Dresden, Germany\\
    \texttt{\{muah.kim, rick.fritschek, rafael.schaefer\}@tu-dresden.de}
	\end{tabular}
    }
}
\maketitle

\begin{abstract}
It is a known problem that deep-learning-based end-to-end (E2E) channel coding systems depend on a known and differentiable channel model, due to the learning process and based on the gradient-descent optimization methods. This places the challenge to approximate or generate the channel or its derivative from samples generated by pilot signaling in real-world scenarios. Currently, there are two prevalent methods to solve this problem. One is to generate the channel via a generative adversarial network (GAN), and the other is to, in essence, approximate the gradient via reinforcement learning methods. Other methods include using score-based methods, variational autoencoders, or mutual-information-based methods. In this paper, we focus on generative models and, in particular, on a new promising method called diffusion models, which have shown a higher quality of generation in image-based tasks. We will show that diffusion models can be used in wireless E2E scenarios and that they work as good as Wasserstein GANs while having a more stable training procedure and a better generalization ability in testing. 
\end{abstract}


\section{Introduction}

\subsection{DL-Based E2E Channel Coding Optimization} 
In a conventional communication system, practical codes, i.e., encoders and decoders, need careful adaption to the specific channel setup, to achieve the maximal transmission rate for reliable communication. This can be a demanding task in dynamically changing environments and would ultimately demand a more automated approach towards this problem, especially with non-standard channel models. 
To fill this gap, many studies introduced deep learning (DL) for optimizing not only encoders and decoders but also modulators and demodulators separately until \cite{o2016learning, o2017introduction} proposed an end-to-end (E2E) framework that jointly optimizes all of those blocks.
The E2E framework resembles the autoencoder (AE) model with a wireless channel block in between the encoder and the decoder parts\cite{o2016learning}, whose encoder part includes a modulator and spreads the encoded and modulated symbols to combat the channel noise instead of compressing the input as AEs normally do. Similarly, the decoder and the demodulator are combined in one neural network (NN) model.
Surprisingly, AE-based E2E communication frameworks have shown a strong error probability performance on par with classically-designed channel coding methods or better than them in some cases \cite{jiang2019turbo, ProductAE}.

To train the encoder by using gradient-based optimizers in DL, the channel connecting the encoder and the decoder needs to be known and differentiable to apply the back-propagation algorithm. This is not guaranteed in general because channels are often hard to be described by a simple mathematical model, and the model is not necessarily differentiable. As a solution, there are studies: to generate the channel by a generative model \cite{ye2018channel,dorner2020wgan,GAN-OShea} and to use reinforcement learning (RL) and a policy gradient method \cite{RL-Paper}, which essentially approximates the gradient of the channel, by perturbing the channel probing variable and comparing the change in rewards.

In this paper, we follow the approach using generative models and use a new generative method called diffusion models, which diffuse data samples by adding noise until they follow a normal distribution and learn to denoise the noisy samples to reconstruct the data samples. New samples can now be generated, by applying this procedure to generated normal distribution samples.
Diffusion models have drawn huge attention in computer vision due to their impressive capability of generating diverse examples with details. There are mainly three kinds of diffusion models with distinct ways of perturbing the sample and learning to denoise \cite{croitoru2022diffusion}, called diffusion-denoising probabilistic models (DDPMs) \cite{sohl2015deep, ho2020denoising}, noise-conditioned score networks (NNCNs) \cite{song2019generative}, and models using stochastic differential equations (SDEs) \cite{song2020score}. This paper uses a DDPM to generate a channel distribution since there have been many studies using DDPM for a conditional generation, and a channel is typically defined by a conditional distribution.

\begin{figure*}[t]
\centering
\begin{subfigure}{\textwidth}
\centering
\includegraphics[width = \textwidth]{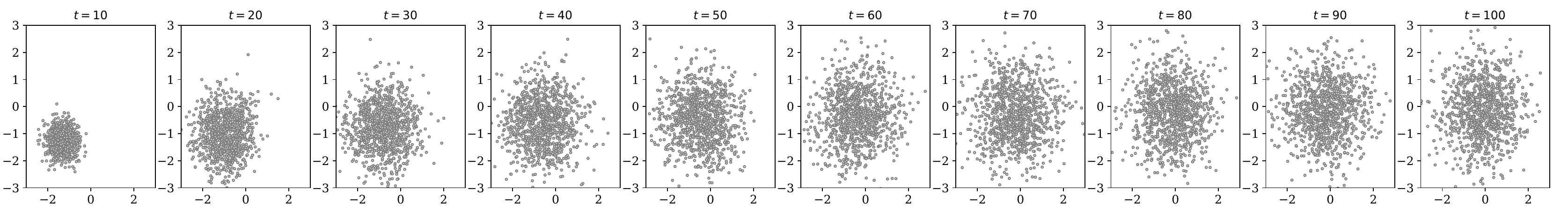}   
\subcaption{Diffusion process.}\label{sfig:diff}
\end{subfigure}
\begin{subfigure}{\textwidth}
\centering
\includegraphics[width = \textwidth]{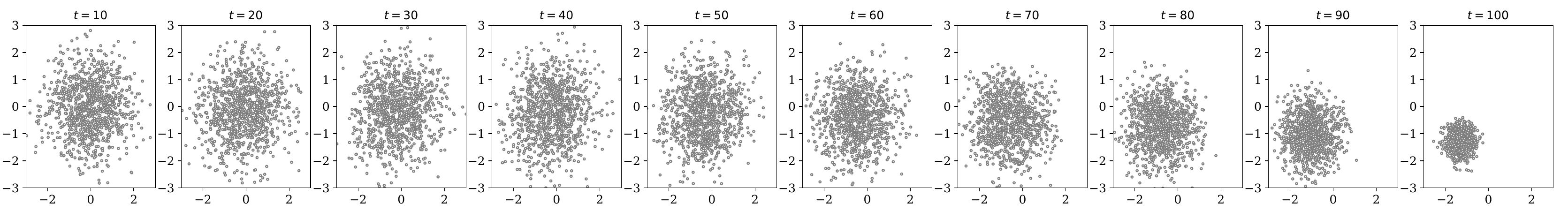}
\subcaption{Denoising process.}\label{sfig:deno}
\end{subfigure}
\caption{Snapshots of diffused samples and denoised samples for an example with $T=100$, $\beta_t=0.05$ and a sample distribution $\mathcal{N}(-1.32, 0.28^2)$ for both dimensions. For denoising, the same model explained in subsection \ref{ssec:AWGN_channel} was used. \vspace{-0.5cm}}\label{fig:diff_deno}
\end{figure*}
\subsection{Related Works: E2E Learning with Generated Channel}
Along with the development of machine learning, there have been active studies about channel estimation by using generative networks and AE-based joint optimization of channel coding and modulation. This subsection only considers the papers studying E2E communication performance by using channel generation with a generative model. This approach was first evaluated in \cite{ye2018channel} with a conditional GAN by the symbol error rate (SER) of the whole framework for an additive white Gaussian noise (AWGN) channel and a Rayleigh fading channel in comparison to the traditional encoding and modulation methods and to the AE trained with the channels with coherent detection. The $E_b/N_0$ vs. SER graph shows that the AE-based coding works almost the same as the traditional method for those channel models, but the generated channel results in an apparent gap, compared to the framework with coherent detection. This approach was further advanced by using a Wasserstein GAN \cite{dorner2020wgan}, which shows a strong channel estimation performance for an actual over-the-air channel and a comparable bit error rate (BER) as good as an RL-based framework under an AWGN channel. A residual-aided GAN \cite{jiang2022residual} was proposed to stabilize the training process, and it shows not only an almost-optimal SER performance but also lower training time compared to WGAN. However, GAN-based approaches suffer from mode collapse when the channel distributions are more sophisticated, for instance, \cite{dorner2020wgan} mentioned a WGAN suffered from mode collapse when it was trained for a channel with 5 random taps.

This led to studies of using other generative networks, which have shown better mode coverage in computer vision task, such as normalizing flows \cite{weisser2021generative}, score-based generative model \cite{arvinte2022score}, variational AE \cite{xia2022generative}, for better performance under more complicated channels. These studies focus on channel estimation performance and have not provided an E2E communication performance yet. To the best of our knowledge, a diffusion model has not been employed for channel estimation in the literature. This paper introduces diffusion models for channel generation by proposing a diffusion model conditioned on the message information. With the synthetic channels produced by the diffusion model, we test the E2E framework's SER performance under an AWGN channel and a Rayleigh fading channel and compare the results to an E2E framework with the defined channel model and another framework trained with a synthetic channel generated by a WGAN. The experimental results show that the proposed model achieves SER almost the same as the channel-aware case and better generalization ability to the high $E_b/N_0$ region compared to WGAN.

\section{Diffusion Models}\label{ssec:diff_model}
A diffusion model consists of a forward and a reverse diffusion process. Given an input sample from some data distribution $x_0\sim q(x)$, we can define the forward diffusion process by adding Gaussian noise to the sample over $T$ time steps, producing increasingly noisy samples $x_1,\ldots,x_T$ of the original sample. This process can be defined with
\begin{align}
    q(x_{1:T}|x_0)&=\prod_{t=1}^T q(x_t|x_{t-1})\\
    q(x_t|x_{t-1})&\sim\mathcal{N}(x_t; \sqrt{1-\beta_t}x_{t-1}, \beta_t\textbf{I}),\label{diffusion_eqn}
\end{align}
where $\{\beta_t\in(0,1)\}_{t=1}^T$. For $T\!\rightarrow \!\infty$, $x_T$ approaches an isotropic Gaussian distribution with covariance $\Sigma\!=\!\sigma^2\mathbf{I}$ for some $\sigma\!>\!0$. From \eqref{diffusion_eqn}, note that the noisier sample $x_t$ is simply a scaled mean from the previous sample $x_{t-1}$ in the process with additional covariance proportional to $\beta_t$, i.e., for some random vector $z_{t-1}\!\sim\!\mathcal{N}(0,\textbf{I})$ for all $t$ we have
\begin{align}
	x_t = \sqrt{1-\beta_t}&x_{t-1} +\sqrt{\beta_t} z_{t-1}.
\end{align}
With the property of the Gaussian distribution that for $z\sim\mathcal{N}(0,1)$, it holds that $\mu+\sigma z\sim \mathcal{N}(\mu, \sigma^2)$\footnote{Sometimes referred to as reparametrization trick in the ML literature}, and the sum properties of two Gaussian random variables, one can sample $x_t$ directly from $x_0$, through recursively applying these properties to arrive at
\begin{equation}
    q(x_t|x_0)\sim\mathcal{N}(x_t;\sqrt{\bar{\alpha}_t}x_0,(1-\bar{\alpha}_t)\mathbf{I}),\label{xt from x0 and z0}
\end{equation}
where $\alpha_t=1-\beta_t$ and $\bar{\alpha}_t\!=\!\prod_{i=1}^t\alpha_i$, see \cite{luo2022understanding} for details. The trick is now, that if we can reverse the process and sample from $q(x_{t-1}|x_t)$, we can reverse the whole chain and generate data samples from $q(x)$, by sampling from $x_T\!\sim\!\mathcal{N}(0,\mathbf{I})$ and applying the reverse process. Here, one can use that when $\beta_t \ll 1$ for all $t$, the reverse distribution $q(x_{t-1}|x_t)$ has the same functional form as the forward one, therefore it is a Gaussian distribution too \cite{sohl2015deep,feller2015theory}. However, this reverse distribution is not given and should be estimated by learning a model $p_\theta$ with
\begin{align}
    p_{\theta}(x_{0:T})&=p(x_T)\prod_{t=1}^T p_{\theta}(x_{t-1}|x_t)\\
    p_{\theta}(x_{t-1}|x_t) &\sim\mathcal{N}(x_{t-1};\mu_{\theta}(x_t,t),\Sigma_{\theta}(x_t,t)).
\end{align}

The task is now to learn the mean $\mu_\theta(x_t,t)$ and the covariance $\Sigma_\theta(x_t,t)$ by optimizing the weights $\theta$ with DL.

Knowing $q(x_t|x_0)$, and $q(x_{t-1}|x_0)$, by using Bayes rule, one can derive $q(x_{t-1}|x_t,x_0)$ in the same form as
\begin{equation}
    q(x_{t-1}|x_t,x_0)\sim \mathcal{N}(x_{t-1};\mu_{q}(x_t,t),\Sigma_{q}(x_t,t)),
\end{equation}
with
\begin{equation}
    \mu_{q}(x_t,t)=\frac{\sqrt{\alpha_t}(1-\bar{\alpha}_{t-1})}{1-\bar{\alpha}_t}x_t + \frac{\sqrt{\bar{\alpha}_{t-1}}(1-\alpha_t) }{1-\bar{\alpha}_t}x_0
\end{equation}
and
\begin{equation}
    \Sigma_{q}(x_t,t)=\frac{(1-\alpha_t)(1-\bar{\alpha}_{t-1})}{1-\bar{\alpha}_t}\mathbf{I}=\sigma^2_q(t)\mathbf{I}.
\end{equation}
Here, one can see that the covariance only depends on constant terms, and one only needs to learn a parameterized mean function. One can for example now minimize the KL divergence between $q(x_{t-1}|x_t,x_0)$ and $p_{\theta}(x_{t-1}|x_t)$, which leads to 
\begin{equation}
\arg\min_{\theta} \frac{1}{2\sigma^2_q(t)} \frac{\bar{\alpha}_{t-1}(1-\alpha_t)^2}{(1-\bar{\alpha}_t)^2}||\hat{x}_{\theta}(x_t,t)-x_0||_2^2\label{predict x_0},    
\end{equation}
where $\hat{x}_\theta(x_t,t)$ is the predicted $x_0$ by the NN by using the noisy data $x_t$ at time step $t$.
This shows, that learning the denoising process is basically learning the original data from noisy samples, which works as a generative model. Equivalently, one might express $x_0$ by $x_t$ and $z_0$, through \eqref{xt from x0 and z0}, to arrive at
\begin{equation}
    \mu_q(x_t,x_0)=\frac{1}{\sqrt{\alpha}_t}x_t-\frac{1-\alpha_t}{\sqrt{1-\bar{\alpha}_t}\sqrt{\alpha_t}}z_0,
\end{equation}
which one can approximate with 
\begin{equation}
    \mu_{\theta}(x_t,t)=\frac{1}{\sqrt{\alpha}_t}x_t-\frac{1-\alpha_t}{\sqrt{1-\bar{\alpha}_t}\sqrt{\alpha}_t}\hat{z}_{\theta}(x_t,t),
\end{equation}
where $\hat{z}_\theta(x_t,t)$ is the estimated noise in $x_t$ learned by an NN.
This then yields\footnote{Details for this and the previous derivation can be found in \cite{luo2022understanding}.} the following optimization
\begin{equation}
   \arg\min_{\theta} \frac{1}{2\sigma^2_q(t)} \frac{(1-\alpha_t)^2}{(1-\bar{\alpha}_t)\alpha_t}||\hat{z}_{\theta}(x_t,t)-z_0||_2^2.
\end{equation} This is also reasonable, since predicting the noise should be equivalent to \eqref{predict x_0}, if $x_T$ is known. It has been empirically shown in \cite{ho2020denoising} that this optimization works better without the factor in front, which leads to the following loss function 
\begin{align}
\mathcal{L} = \underset{t\sim U[1,T], x_0\sim q(x_0), z_0\sim \mathcal{N}(0,\textbf{I})}{\mathbb{E}} \left[||\hat{z}_{\theta}(x_t,t)-z_0||_2^2\right],\label{eq_loss}
\end{align}
where $t\sim U[1,T]$ means that $t$ is uniformly distributed over $\{1, 2, \dots, T\}$, and we assume that $\Sigma_{\theta}(x_t,t)=\beta_t \textbf{I}$.

Fig. \ref{sfig:diff} and Fig. \ref{sfig:deno} visualize the diffusion and denoising process by showing the samples for every 10 time steps out of $T=100$ in each process. Along the time step, the diffused samples converge to normally distributed samples, whereas the denoised samples approach to the original samples.

\subsection{Conditional Diffusion Models for Channel Estimation}
To use the diffusion model for generating a differentiable channel, two conditions need to be satisfied: 1) the diffusion model should generate a conditional distribution, and 2) the channel input should be trainable parameters of the diffusion model. 
A stochastic channel is characterized by a conditional distribution of the channel output conditioned by the channel input. To generate a channel, the generative model should be capable of generating multiple distributions controlled by a condition $c$. We consider $M$ different conditions and the data sample distributions $q(x_0|c)$ for each condition $c$. For simplicity, we assume the same forward process for all possible conditions, i.e., $\beta_t$ does not depend on $c$. Meanwhile, the backward step works conditionally by $c$ as $p_\theta(x_{t-1}|x_t, c)\sim\mathcal{N}(x_{t-1};\mu_\theta(x_t,t,c), \Sigma_\theta(x_t,t,c))$ with the same distribution but with distinct mean and covariance.
With the same choice of the simplified loss as in the general case, the loss of the conditional diffusion model becomes
\begin{align}
\mathcal{L}_c =\underset{t\sim U[1,T], c \sim p(c),x_0\sim q(x_0|c), z_0\sim \mathcal{N}(0,\textbf{I})}{\mathbb{E}}\left[||\hat{z}_{\theta}(x_t,t|c)\!
-\!z_0||_2^2\right]\label{eq_loss_c},
\end{align}
where $\hat{z}_\theta(x_t,t|c)$ is defined to satisfy 
\begin{align}
\mu_\theta(x_t,t,c) = \frac{1}{\sqrt{\alpha_t}} x_t - \frac{1-\alpha_t}{\sqrt{1-\bar{\alpha}_t}\sqrt{\alpha_t}}\hat{z}_\theta(x_t,t|c) .\label{eq_mu_theta_c}
\end{align}
The random vector $\hat{z}_\theta(x_t,t|c)$ can be learned by using DL.

For channel estimation, we want to learn the noisy codeword $y=x_0$ as our data point, conditioned by the channel input $c$,  from a noise distribution $x_T$.
For training the AE with the generated channel, one can simply plug in the NN model for $\hat{z}_\theta(x_t,t|c)$ into the process, which connects the channel input $c$ as the conditioning variable with the channel output, as the generated data point. Also note that the backward step $p_\theta(x_{t-1}|x_t,c)$ of the diffusion model follows a normal distribution whose mean satisfies \eqref{eq_mu_theta_c}, which only contains scaling and addition of $x_t$ and $\hat{z}_\theta(x_t,t|c)$. With this, the chain rule can be applied without problems when the partial derivative of the channel output with respect to the channel input is obtained for training the AE model, i.e., for back-propagation. Thus, the channel becomes differentiable when the NN model for $\hat{z}_\theta(x_t,t|c)$ uses the channel input as trainable parameters. 

\begin{figure*}[t!]
    \centering
    \begin{subfigure}{0.445\columnwidth}
        \centering
        \includegraphics[width=\columnwidth]{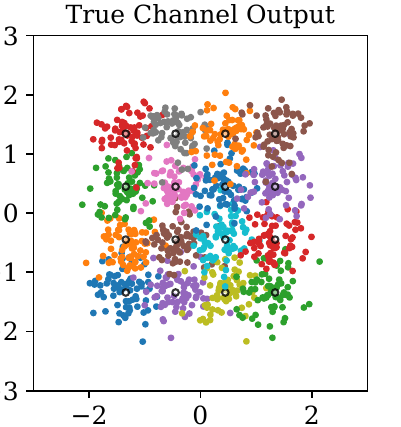}
    \end{subfigure}    
    \begin{subfigure}{0.445\columnwidth}
        \centering
        \includegraphics[width=\columnwidth]{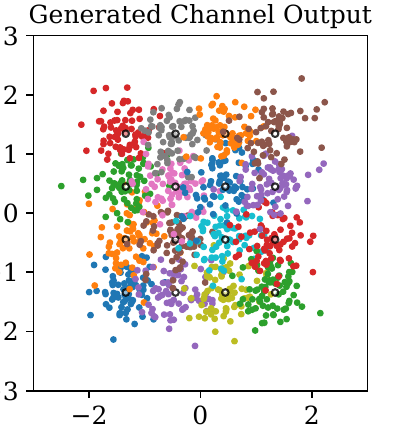}
    \end{subfigure}
    \begin{subfigure}{0.56\columnwidth}
        \centering
        \includegraphics[width=\columnwidth]{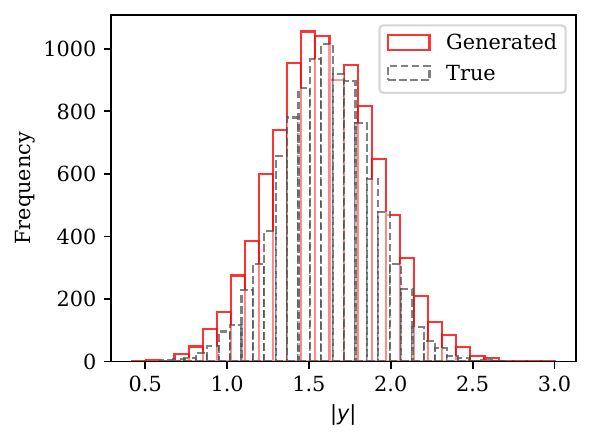}
    \end{subfigure}    
    \begin{subfigure}{0.52\columnwidth}
        \centering
        \includegraphics[width=\columnwidth]{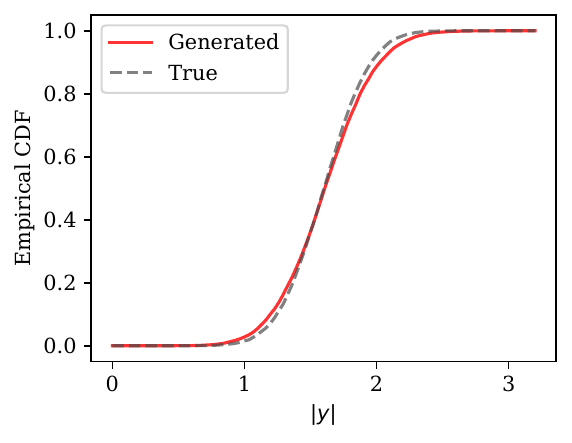}
    \end{subfigure}
    \caption{The generative performance of the diffusion model is tested for an AWGN channel with 16-QAM modulated symbols and $E_b/N_0=\qty{5}{\decibel}$. The constellation of channel output is drawn with 70 samples for each message. For m=3, the histogram and empirical CDF of the channel output's norm are obtained with $10^4$ samples and compared.}\label{fig:Constell_AWGN_QAM16}
\end{figure*}

\section{End-to-end System Model and Algorithm}
\subsection{Autoencoder and Conditional Diffusion Models}\label{ssec:AE_cDiff}
Consider an AE model, where a generated channel is embedded in between the encoder and the decoder parts instead of the real channel. The encoder $f_{\theta}$ maps each one-hot-encoded message, labeled by $m\in\{1,2,\dots,M\}$, to a codeword defined in an $n$-dimensional real vector space, i.e., $f_{\theta}(m)\in\mathbb{R}^n$. We use an NN whose input and output layers have $M$ and $n$ nodes, respectively, with some hidden layers in between. The encoder model's output is normalized to prevent diverging and meet the power constraint. Denote the generated channel output as $g(f_{\theta}(m))$ which also lies in $\mathbb{R}^n$. The decoder $h_{\theta}$ has the mirrored structure as the encoder, taking $n$-dimensional real vectors and producing $M$ values indicating the predicted likelihood of each message, i.e., $h_{\theta}\left( g(f_{\theta}(m)) \right) \in (\mathbb{R}^+)^m$. 

The conditional diffusion model learns the noise component $\hat{z}_\theta(x_t,t|c)$ in $x_t$, which behaves differently for each condition $c$. We follow the architecture proposed by \cite{ho2020denoising} as a baseline and modify it for channel generation. A diffusion model of \cite{ho2020denoising} takes $x_t$ as an input and returns the noise component at every $t$. Besides $x_t$, we feed the diffusion model the encoded channel input $f_\theta(m)$ as an input to make the channel differentiable. Hence, we define the input layer with $2n$ nodes, each of which is for $x_n$ and $f_\theta(m)$. Instead of training $T$ distinct models for each time step, a diffusion model uses one NN model whose hidden layers are conditioned on $t$ by multiplying embeddings of $t$ \cite{DDPMGitHub}. Furthermore, we condition the parameters of the model on $c$ by multiplying embeddings of message label $m$. In summary, the condition is defined by $c\coloneqq (m,f_\theta(m))$, where $m$ is used to condition the hidden layers of the model, and $f_\theta(m)$ is fed to the model as an additional input.

Note that the output of the diffusion model is not the same as the generated channel. The samples following the channel output distribution can be generated by calculating $x_0(c)$ by using the following equation 
\begin{align}
&x_{t-1}(x_t,t,c) = \mu_\theta(x_t, t, c) + (1-\alpha_t) z_0\\
&= \frac{1}{\sqrt{\alpha_t}} x_t - \frac{1-\alpha_t}{\sqrt{1-\bar{\alpha}_t}\sqrt{\alpha_t}}\hat{z}_\theta(x_t,t|c) + (1-\alpha_t) z_0,
\end{align}
for $t=\{T, T-1, \dots, 1\}$ with the learned $\hat{z}_\theta(x_t,t|c)$ where $x_T$ and $z_0$ are samples of the standard normal distribution. 

The AE model and the diffusion model are trained iteratively by fixing the other one when one of them is trained because they have distinct objective functions and their optimization depends on each other. The AE model is trained to minimize the cross entropy of the one-hot-encoded message and the decoder output, i.e.,
\begin{align}
\mathcal{L}_{AE}\! =\! \mathbb{E}_{m\sim p(m)}\left[ -\sum_{i=1}^M\mathbbm{1}_{i=m}\log \left( h_\theta\left(g\left(f_\theta(m)\right)\right) \right)_i \right],
\end{align}
where $\mathbbm{1}$ is the indicator function. During the training of the AE, the generated channel function $g$ is fixed and only the parameters of the AE are optimized. Similarly, when the diffusion model is trained by using the loss function \eqref{eq_loss_c}, only the parameters in the diffusion model are optimized with the encoder fixed, which determines $f_\theta(m)$ in $c$. This alternating training is repeated until the models converge. During the early stage, the diffusion model is trained with a large number of epochs for each iteration, and we decreased it once the models converge. The encoder changes much in the early stage, so we allow a smaller number of epochs for each iteration so that the generative model is up-to-date with the encoder. When the constellation of the encoder does not change much by iterations due to convergence, we allow the AE a larger number of epochs for fine-tuning.

\section{Simulation Results}
We evaluate the channel generation and the E2E communication performance of the diffusion model under an AWGN channel and a real Rayleigh fading channel. For all experiments, the messages are assumed to be uniformly distributed. The AEs and diffusion models follow the structure explained in subsection \ref{ssec:AE_cDiff}, but those for each channel model are allowed to have a distinct depth, activation functions, and the number of hidden neurons.

\begin{figure}[t]
    \centering
    \includegraphics[width=0.43\textwidth]{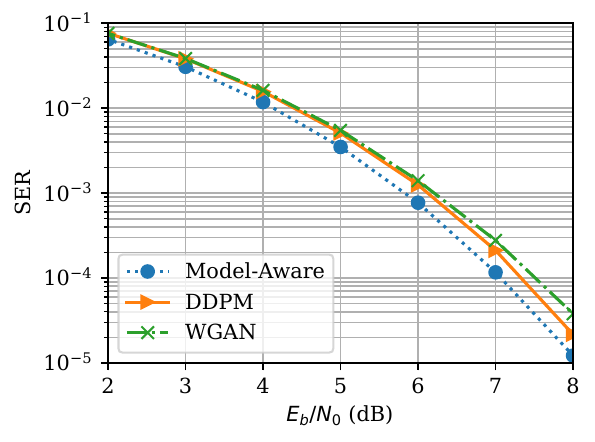}
    \caption{The E2E SER for varying $E_b/N_0$ values is tested for an AWGN channel and generated channels. \vspace{-0.5cm}}
    \label{fig:SER_AWGN}
\end{figure}

\subsection{Channel Generation: 16-QAM for AWGN Channel}\label{ssec:ch_gen_test}
This subsection demonstrates the conditional diffusion model's channel generative performance by using an experiment with a 16-ary quadrature amplitude modulation (QAM) and an AWGN channel with $E_b/N_0=\qty{5}{\decibel}$.
The AWGN channel is defined as $Y^n = X^n + Z^n$, where $X^n$ is the channel input and $Z^n$ is a channel noise vector independent of $X^n$ such that $Z^n\sim \mathcal{N}(0,\sigma^2\textbf{I})$ for some $\sigma>0$. For channel input, we used symbols modulated by the normalized 16-QAM and the corresponding $\sigma$ such that $E_b/N_0=\qty{5}{\decibel}$.
For the diffusion process, we set $T=100$ and $\beta_t=0.05$ for all $t=1,2,\dots,100$ for simplicity. And we used the conditional diffusion model with 3 hidden layers, each of which has 64 neurons conditioned on $t$ and $m$ and activated by the Softplus function. The output layer is a simple linear layer. The diffusion model was optimized by using an accelerated adaptive moment estimation (Adam) optimizer with a learning rate starting from $10^{-3}$ and decreasing to $10^{-4}$. Fig.~\ref{fig:Constell_AWGN_QAM16} visualizes the constellations of the channel output for the real channel and the generated channel in the first two subfigures from the left, which show very similar distributions. The conditioning capability can be proved by the channel output distributions generated for each message. To check the distribution more precisely, the histogram and empirical cumulative density function (CDF) of the channel output's norm were obtained with $10^4$ samples for each message. Due to space limits, only the graphs of $m=3$ are shown in Fig.~\ref{fig:Constell_AWGN_QAM16}. Both graphs demonstrate the generated distribution is almost identical to the true distribution, and similar results were found for the other messages too.

\subsection{E2E Learning: AWGN Channel}\label{ssec:AWGN_channel}
 For the experiment of E2E learning with an AWGN channel, we used $n=7$, $M=16$ and trained the models with $E_b/N_0=\qty{5}{\decibel}$. With the AE model, the channel input is set by the encoder's output as $X^n=f_\theta(m)$, and the channel output $Y^n$ is decoded as $h_\theta(Y^n)$.
 Since an AWGN channel is a simple channel, the encoder and decoder were defined by simple NN models with two hidden linear layers with $M=16$ neurons activated by exponential linear unit (ELU) activation. The encoder and the decoder were trained simultaneously by using a Nesterov-accelerated adaptive moment estimation (NAdam) optimizer with a learning rate of $10^{-3}$.
 For diffusion, we set $T=50$ and $\beta_t=0.05$ for all $t$ for simplicity. The diffusion model architecture and the optimizer for it were defined the same as the previous subsection except for the learning rate, which we used down to $10^{-5}$ for convergence. For both the AE model and the diffusion model, we used a randomly generated message dataset of $10^5$ samples with a batch size of $3000$.    

The E2E SER was evaluated for integer $E_b/N_0$ values from $\qtyrange{2}{8}{\decibel}$ and illustrated in Fig.~\ref{fig:SER_AWGN} with a label of DDPM. Besides the proposed framework, we implemented two other E2E frameworks, whose AEs are trained with the defined channel model, labeled as Model-Aware, and a generated channel by using a WGAN.
For the Model-Aware framework, the simulated channel output was directly used for training the AE. The SER curve of it can be regarded as the target value that we want to achieve with the generative models.  
For the generator and the critic of the WGAN, we used NNs with two linear hidden layers with 128 nodes followed by the rectified linear unit (ReLU) activation. To make the models differential, we feed the channel input to the input layer of both the generator and the critic models. For training the models, the learning rate of $10^{-4}$ was used with root mean square propagation (RMSprop) optimizer. 

 The SER values decrease along with $E_b/N_0$, drawing concave curves for all cases. The diffusion model and the WGAN model both show the SER values close to the target curve, whereas the diffusion model's curve gives a slightly lower value roughly from $E_b/N_0$ values larger than \qty{6}{\decibel}.

\begin{figure}
    \centering
    \includegraphics[width=0.43\textwidth]{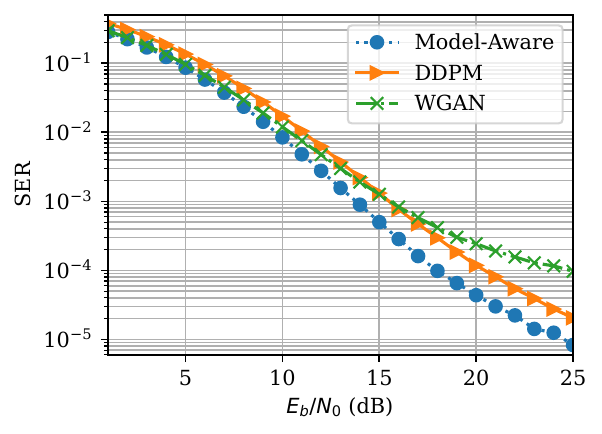}
    \caption{The E2E SER for varying $E_b/N_0$ values is tested for a Rayleigh fading channel and generated channels.\vspace{-0.5cm}}
    \label{fig:SER_Rayleigh}
\end{figure}

\subsection{E2E Learning: Rayleigh Fading Channel}
 The real Rayleigh fading channel is defined as $Y^n = H^n\circ X^n + Z^n$, where $\circ$ is the element-wise multiplication operator and $H^n=(H_1, H_2, \dots, H_n)$ satisfies $\Pr(H_i=x)=\frac{x}{\sigma_R^2}\exp(-x^2/(2\sigma_R^2))$ for all $i=\{1,2,\dots,n\}$ and for some $\sigma_R>0$. The additive noise $Z^n$ is identically defined as in subsection \ref{ssec:ch_gen_test}. We assumed a real Rayleigh fading channel with $n=7$, $\sigma_R=1$ for $M=16$ messages and trained the models with $E_b/N_0=\qty{12}{\decibel}$. We used the same architecture of the diffusion model and the AE model as those used for an AWGN channel in subsection \ref{ssec:AWGN_channel}. 

 Fig.~\ref{fig:SER_Rayleigh} illustrates the test SER values of the proposed E2E framework with a diffusion model with a comparison to Model-Aware and WGAN frameworks. The WGAN model was similarly defined except for having 256 neurons in each hidden layer and the learning rate set by $\num{5e-5}$. The trained models are tested for integer $E_b/N_0$ values from \qtyrange{1}{25}{\decibel}. 
 The SER curves of DDPM and Model-Aware frameworks are decreasing linearly in the log scale plot, whereas the WGAN curve is decreasing along with them and starting to diverge around \qty{17}{\decibel}. Note that the WGAN and the DDPM show almost the same performance in the low $E_b/N_0$ region, whereas the DDPM outperforms WGAN in the high $E_b/N_0$ region and keeps decreasing in parallel to the target curve. This can be interpreted as the diffusion model gives a better generalization to other $E_b/N_0$ values.

\subsection{Discussion and Future Work}
The experimental results prove that the diffusion model can learn the channel distribution with high precision and also that the E2E framework with it achieves an SER almost close to that of the channel-aware framework for an AWGN channel and a real Rayleigh fading channel. In both experiments with those channel models, the diffusion model showed strong generalization performance in the high $E_b/N_0$ region, outperforming WGAN that showed a saturated behavior for $E_b/N_0$ values that are higher than the value used for training. 
This indicates the potential of the diffusion model as an alternative to WGAN-based channel generation, especially for the channels where WGAN does not work stably. For example, \cite{dorner2020wgan} mentions that the WGAN model was stuck in a single mode when it was used for learning channels with 5 random taps. In computer vision, the diffusion model has shown better mode coverage than GANs, so it could be a possible solution to overcome the difficulty that WGAN has. As a future work, we consider exploring more realistic channel models and improving the architecture and algorithm of the diffusion models. Plus, the sample complexity and training time analysis can be studied for a fairer comparison.
\section{Conclusion}
We investigated the AE-based E2E communication framework with a synthetic channel generated by a conditional diffusion model. The strong channel estimation performance of the diffusion model is demonstrated by using an AWGN channel with a 16-QAM. And then E2E learning capability of it is evaluated by $E_b/N_0$ vs. SER curves for an AWGN channel and a real Rayleigh fading channel with a comparison to a WGAN. Under both channel models, both the diffusion model and the WGAN model achieved SER values close to the framework trained with the real channel in the $E_b/N_0$ region around the value used for training. However, the diffusion model provided an SER curve constantly decreasing in parallel to the target curve, whereas the SER curve of WGAN starts diverging in the high $E_b/N_0$ region. The framework with the diffusion model seems to generalize the error-correcting performance to the high $E_b/N_0$ region better than a WGAN. This signals a possibility that a diffusion model can be a solution for modeling channels that require stable convergence and better generalization capability. 
\newpage
\bibliographystyle{IEEEbib}
\bibliography{refs}
\end{document}